# A dislocation-based explanation of quasi-elastic release in shock-loaded aluminum

Song-Lin Yao, Xiao-Yang Pei, Ji-Dong Yu[a], Jing-Song Bai, and Qiang Wu

*National Key Laboratory of Shock Wave and Detonation Physics, Institute of Fluid Physics, China Academy of Engineering Physics, Mianyang, Sichuan 621900, China*

## Abstract

A novel explanation of the quasi-elastic release phenomenon in shock-compressed aluminum is presented. A dislocation-based model, taking into account dislocation substructures and evolution, is applied to simulate the elastic–plastic response of both single-crystal and polycrystalline aluminum. The calculated results are in good agreement with experimental results from not only the velocity profiles but also the shear strength and dislocation density, which demonstrates the accuracy of our simulations. Simulated results indicate that dislocation immobilization during dynamic deformation results in a smooth increase of yield stress, which leads to the quasi-elastic release, while the generation of dislocations caused by the plastic release wave results in the appearance of a transition point between the quasi-elastic release and the plastic release in the profile.

## I. INTRODUCTION

Plate-impact experiments have been applied as a routine method to quantify the response of solids to high stresses and high strain rates for several decades.[1,2] Measurements of loading and unloading wave profiles in plate-impact experiments provide the information needed to estimate the yield stress and the shear strength that are critical in high-pressure constitutive model studies.[3,4,5] The typical wave profile and schematic view of wave propagation in shock-loading experiments are presented in Figure 1. For ideal elastic–plastic materials, a distinct two-wave structure is expected (e.g., see the red line in Figure 1). For realistic materials, however, such a two-wave structure is absent. Instead, the wave profile shows a smooth transition from elastic release to plastic release, as shown by the black line in Figure 1. This phenomenon is termed quasi-elastic release.

Quasi-elastic release reveals the elastic–plastic property of shock-compressed solids. A comprehensive understanding of the physical mechanism of quasi-elastic release is necessary to understand the plastic response of shock-compressed solids and to establish a physics-based constitutive model at high pressure.[6]

---

[a] yujidong@caep.cn

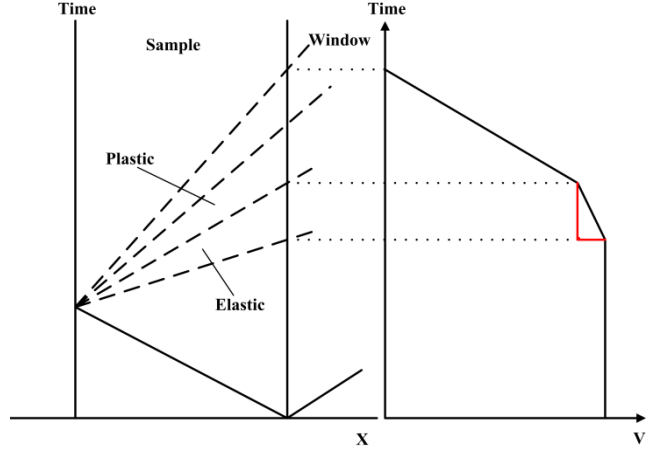
FIG. 1. *x-t* diagram and schematic view of particle velocity history.

Several works have been published to explain the measured quasi-elastic release phenomena. Asay and his coworker proposed an analytical method[4,5] based on the assumption that the material stress state is not on the yield surface but demonstrates a stress distribution with an average equal to the applied stress. Johnson et al. interpreted the quasi-elastic release effect with an internal stress model.[7] In particular, they assumed that release of the back stresses associated with pinned dislocation loops causes an unloading response that is initially elastic but progressively becomes plastic. A two-dimensional meso-scale model was proposed by Dwivedi at al.,[8] who attributed the quasi-elastic response to grain-scale heterogeneities.

Although all these models could predict the quasi-elastic response very well, none considered the basic plastic deformation mechanism, namely, the evolution of dislocations, explicitly. As the plastic shear unit, dislocations play a pivotal role in plastic deformation of crystalline metals.[9] In particular, dislocation slip leads to energy relaxation, while dislocation generation hardens the material. The quasi-elastic response, like other plastic deformations, should also be attributed to dislocation evolution. Therefore, in this paper we aim to calculate the dynamic deformation with dislocation evolution included and to give a more physical interpretation of the quasi-elastic behavior.

In this work, the dynamic response of both single-crystal and polycrystalline aluminum under shock loading is simulated using a viscous constitutive model. The results suggest that dislocation immobilization results in quasi-elastic release, with an unloading speed proportional to the immobilization rate.

## II. MATHEMATICAL MODEL

The dynamic response of materials is described by the continuum mechanics equations in Lagrangian coordinates as follows:

$$\frac{1}{\rho}\frac{d\rho}{dt} = -\sum_{k=1}^{N}\frac{\partial v_k}{\partial x_k}, \qquad (1)$$

$$\rho\frac{dv_i}{dt} = \sum_{k=1}^{N}\frac{\partial \sigma_{ik}}{\partial x_k}, \qquad (2)$$

$$\rho\frac{dU}{dt} = \sum_{k=1}^{N}(-p\frac{\partial v_k}{\partial x_k} + \sum_{i=1}^{N} S_{ik}\frac{d\gamma_{ik}}{dt}), \qquad (3)$$

$$\sigma_{ij} = -p\delta_{ij} + S_{ij}., \qquad (4)$$

where $\rho$ is the substance density, $v_i$ is the mass velocity, $x_k$ is the current coordinate, $U$ is the specific energy, $\sigma_{ij}$ is the tensor of mechanical stresses, $p$ is the hydrostatic pressure, and $S_{ij}$ is the stress deviator. The hydrostatic pressure is controlled by Grüneisen equation of states (EOS) in which $c_0 = 5.320 \times 10^3$ m/s, $\lambda = 1.338$, and $\gamma = 2.0$,[10] while the stress deviators are determined by Hooke's law:

$$S_{ij} = c_{ijkl}[\varepsilon_{kl} - \frac{1}{3}\delta_{kl}\sum_{n=1}^{3}\varepsilon_{nn} - \gamma_{kl}], \qquad (5)$$

where $\varepsilon_{kl}$ is the strain tensor, $\gamma_{kl}$ is the plastic strain tensor, and $c_{ijkl}$ are the elastic constants, which are proportional to hydrostatic pressure:

$$c_{ijkl} = c_{ijkl}^0 + \frac{\partial c_{ijkl}}{\partial p} \times p. \qquad (6)$$

For an isotropic material, Hooke's law can be written as

$$S_{ij} = 2G[\varepsilon_{ij} - \frac{1}{3}\delta_{ij}\sum_{n=1}^{3}\varepsilon_{nn} - \gamma_{ij}], \qquad (7)$$

where $G$ is the shear modulus, which is also proportional to hydrostatic pressure:

$$G = G_0 + \frac{\partial G}{\partial p}p. \qquad (8)$$

The elastic constants of aluminum are listed in Table I.

Table I. Elastic constants. (The units of the shear modulus and elastic constants are GPa.)

| Parameter | $G_0$ | $\partial G/\partial p$ | $c_{11}$ | $c_{12}$ | $c_{44}$ | $\partial c_{11}/\partial p$ | $\partial c_{12}/\partial p$ | $\partial c_{44}/\partial p$ |
|---|---|---|---|---|---|---|---|---|
| Value | 27.8[a] | 1.741[a] | 103.72[b] | 57.52[b] | 28.32[b] | 7.34[b] | 4.05[b] | 2.39[b] |

[a]The shear modulus and its pressure derivative are from experiments(Ref. 11).

[b]Elastic constants and their pressure derivatives are from experiments(Ref. 12).

Dislocations are driven by external applied stress. The governing equation for the dislocation velocity takes the following form:[13]

$$m\frac{dV_D^\beta}{dt} = b(F_D^\beta/b - Y/2) - BV_D^\beta, \qquad (9)$$

where $m$ is the rest mass of a dislocation per unit length; $b$ is the Burgers vector; $F_D^\beta/b = \sum_{ij}S_{ij}n_ib_j/b$ is the resolved shear stress (RSS), i.e., the projection of the applied stress along the slip direction; $B$ is the viscous damping coefficient;[14] $Y/2$ is the critical resolved shear stress (CRSS) below which dislocations are at rest; and $\beta$ denotes the slip direction.

Dislocations are classified as mobile dislocations and immobile dislocations by their role in plastic deformation. Mobile dislocations control plastic slip and some mobile dislocations would be immobilized when interacting with obstacles, while immobile dislocations control the yield strength.

The kinetic equations for the two types of dislocation are written as follows:[15]

$$\frac{d\rho_D^\beta}{dt} = Q_D^\beta - Q_I^\beta - k_ab|V_D^\beta|\rho_D^\beta(2\rho_D^\beta + \rho_I^\beta), \qquad (10)$$

$$\frac{d\rho_I^\beta}{dt} = Q_I^\beta - k_ab|V_D^\beta|\rho_D^\beta\rho_I^\beta, \qquad (11)$$

where $\rho_D^\beta$ and $\rho_I^\beta$ refer to mobile and immobile dislocation density, respectively. Here $Q_D^\beta$ is the generation rate of mobile dislocations and is given by

$$Q_D^\beta = k\left[\frac{BV_D^\beta}{\sqrt{1-(V_D^\beta/c_t)^2}} + Y/2\right]\rho_D^\beta b\left|V_D^\beta\right|, \quad (12)$$

where $k$ refers to the generation coefficient; $Q_I^\beta$ is the immobilization rate and is given by

$$Q_I^\beta = V_I(\rho_D^\beta - \rho_{min})\sqrt{\rho_I^\beta}, \quad (13)$$

where $V_I$ is the characteristic velocity of the dislocations movement during the process of consolidation and $\rho_{min}$ is the minimum dislocation density, which is also the initial dislocation density.

Knowledge of the mobile dislocation density and velocity allows one to calculate the plastic strain rate as

$$\frac{d\gamma_{ij}}{dt} = \frac{1}{2}\sum_\beta (b_i^\beta n_j^\beta + b_j^\beta n_i^\beta)V_D^\beta \rho_D^\beta. \quad (14)$$

The yield strength is determined by the following relation:

$$Y = Y_I + A_I Gb\sqrt{\rho_I}, \quad (15)$$

where $\rho_I$ is the total density of immobile dislocations:

$$\rho_I = \sum_\beta \rho_I^\beta. \quad (16)$$

Table II. Parameters of the plasticity model.

| Parameter | $Y_0$ (MPa) | $b$ (nm) | $k_a$ | $A_I$ | $V_I$ (m/s) | $k$ ($10^{15}$J$^{-1}$) |
|---|---|---|---|---|---|---|
| Value | 22.0[a] | 0.255[a] | 10.0[b] | 10.0 | 7.5 | 6.0 |

[a]These parameters are the same as those used in Krasnikov and his coworkers' model(Ref. 16).

[b]This parameter is the same as that used in Mayer and his coworkers' model(Ref. 15).

The values of parameters of the plasticity model are listed in Table II. Optimal values of adjustable parameters such as $A_I$, $V_I$, and $k$ are obtained by fitting the experimental results. The generation coefficient $k$ endowed with an energy explanation in Ref. 16 is also treated as an adjustable parameter in this paper.

A continuum plasticity routine was used for the two-dimensional simulations, with the viscous constitutive model appended as a customized subroutine. The spatial discretization is 0.5 μm. The mesh size convergence study is described in the following.

# III. RESULTS AND DISCUSSION

We apply the above-mentioned model to simulate the plate-impact experiments performed with both single-crystal and polycrystalline aluminum by Huang and Asay in 2007.[17] Related experimental parameters are listed in Table III.

Table III. Experiment parameters. RL100, RL111, and RL110 denote experiments performed with single-crystal aluminum oriented along ⟨100⟩, ⟨111⟩, and ⟨110⟩, respectively. RL1050 denotes the experiment performed with polycrystalline 1050 aluminum. In these experiments, the samples were mounted on the projectile and impacted directly against the LiF window.

| Exp. No. | $V_p$ (km/s) | Impactor sample (mm) |
|---|---|---|
| RL100-1 | 1.578 | 3.363 |
| RL100-2 | 2.289 | 3.397 |
| RL111-1 | 1.566 | 3.164 |
| RL111-2 | 2.281 | 3.027 |
| RL110-1 | 2.272 | 3.238 |
| RL110-2 | 1.494 | 3.358 |
| RL1050 | 2.301 | 3.336 |

As indicated in the experiments, the impurities and grain size only influence the ambient strength, but not the dynamic yield strength.[18] Thus these factors are not taken into account in this paper. An assumption is made in this paper that, if we ignore the grain boundary, polycrystalline aluminum could be treated as an isotropic single-crystal material since the anisotropy ratio of single-crystal aluminum, namely, $2c_{44}/(c_{11} - c_{12})$, equals 1.21.[19]

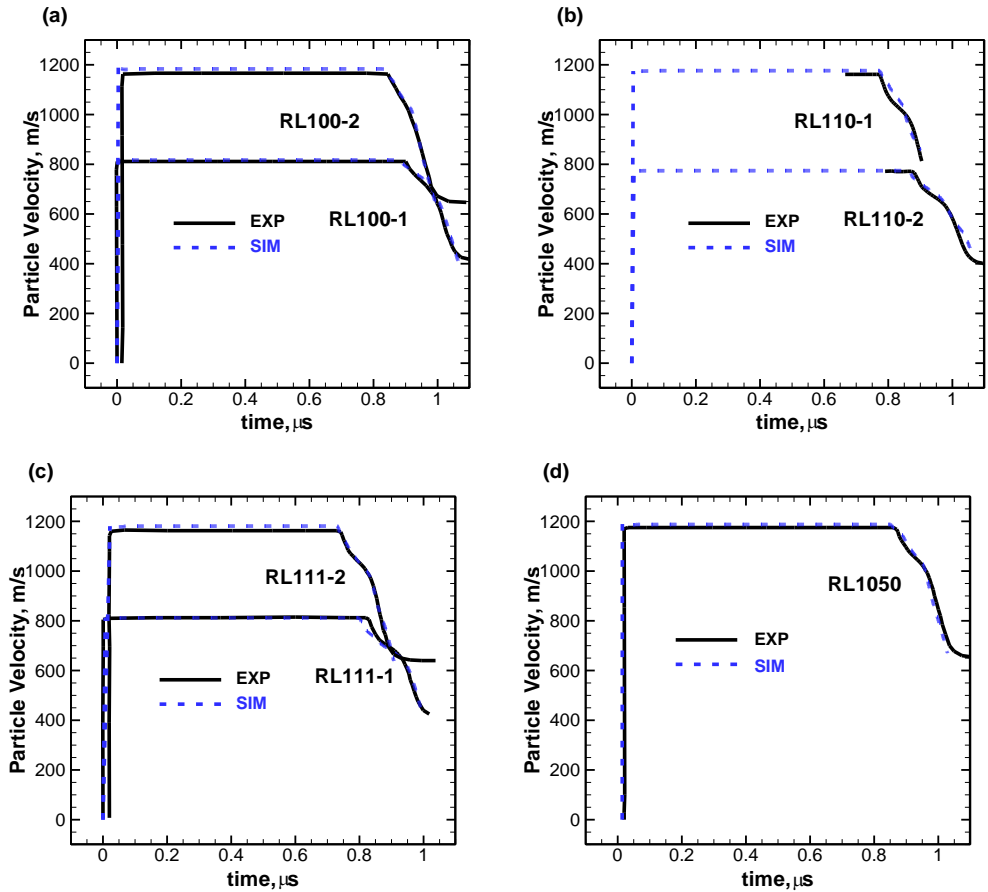

FIG. 2. Comparison of sample/window interface velocity between calculated results and experimental results at different shock directions of single-crystal aluminum (a, b, c) and polycrystalline aluminum (d).

Comparisons between the calculated interface velocities (blue dashed curves) and those obtained from the experiments (black solid curves) of both single-crystal and polycrystalline aluminum are shown in Figure 2. It is obvious that the numerical and experimental results agree well with each other. In particular, the smooth transition from elastic to plastic release is both observed experimentally and well reproduced numerically for all cases. The similar quasi-elastic response detected both in single-crystal and polycrystalline aluminum suggests that microstructural effects other than grain orientation and grain boundaries contribute to the quasi-elastic release behavior. The different number of activated slip planes, which is 8, 6, and initially 4 but progressively becomes 8 out of the possible 12 slip planes for shock propagation along ⟨100⟩, ⟨111⟩, and ⟨110⟩, respectively, leads to the different response under different shock directions.

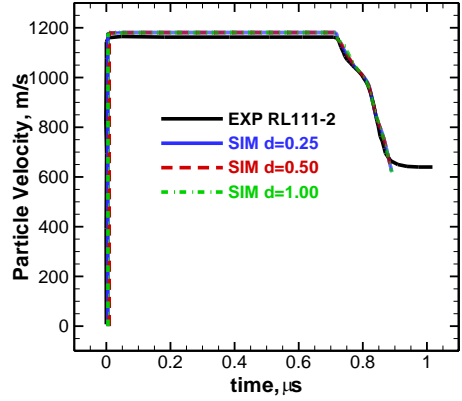

FIG. 3. Calculated velocity profiles of RL111-2 with different mesh sizes.

The calculated results of RL111-2, loaded with a higher velocity, with different mesh sizes are presented in Figure 3. The figure suggests that mesh convergence is good when the mesh size is <1.00 μm.

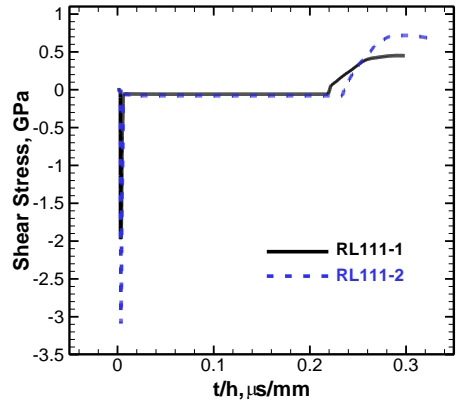

FIG. 4. *In situ* shear stress of experiments when shocking is along 111.

Shear stresses, referred to as the resolved shear stress under uniaxial strain shock loading, of RL111-1 and RL111-2 are presented in Figure 4. When the shock wave reaches a region, the shear stresses increase rapidly to a relative high value and then decrease rapidly to the shear strength owing to plastic relaxation. As the release wave reaches a region and travels through it, the shear stresses are negative then increase toward positive values and approach the reverse shear strength. We define the shear stress at the time when quasi-elastic release begins as the shear strength $\tau_H$, and we define the shear stress at the time when quasi-elastic release ends as the reverse shear strength $\tau_C$. The calculated $\tau_H + \tau_C$ of single-crystal aluminum for different shock directions are compared with experimental results, as displayed in Figure 5. We can learn from the comparison that the calculated results of $\tau_{H}+\tau_{C}$ in all cases but the 100 case match well with experimental results evaluated from release experiments, which demonstrates the capability of the present model in predicting realistic dynamic plastic deformation with adequate accuracy.

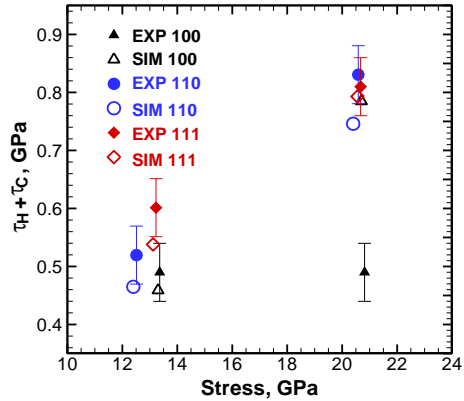

FIG. 5. Comparison of calculated results and experimental results: $\tau_{H}+\tau_{C}$ of single-crystal aluminum for different shock directions.

Figures 6 and 7 show the calculated results for the time history of typical mechanical quantities at a position 79 μm away from the interface in the sample, namely, the time history of RSS and CRSS in Figure 6 and that of mobile and immobile dislocation densities in Figure 7. The sharp rise in CRSS behind the leading edge of the shock front serves as the driving force for dislocation generation and some of the mobile dislocations will be immobilized when interacting with obstacles. After the shock wave, a great many dislocations of both types are generated.

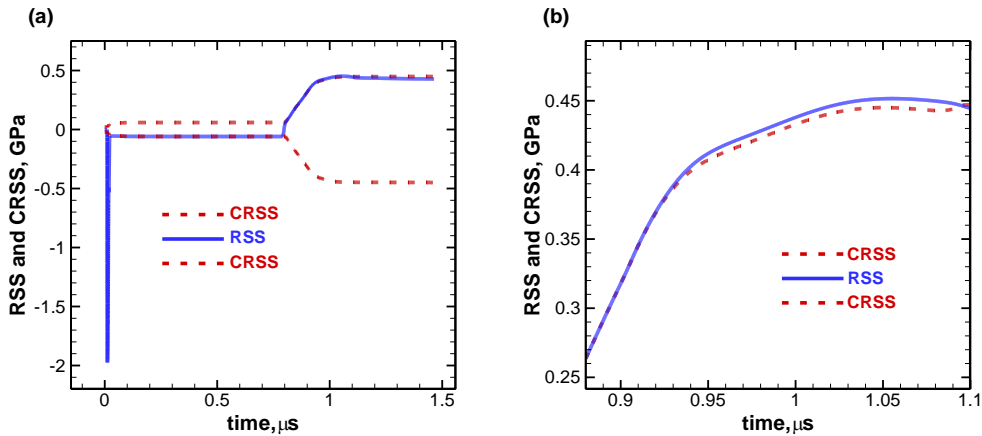

FIG. 6. *In situ* RSS and CRSS of RL111-1.

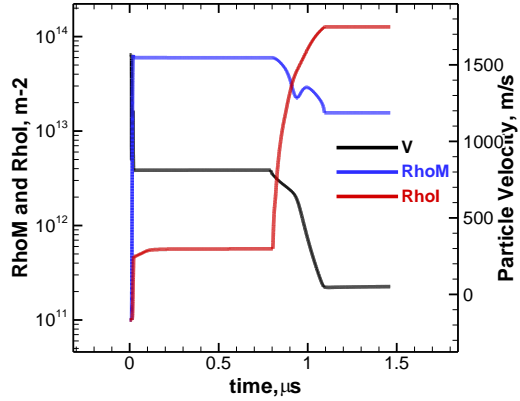

FIG. 7. *In situ* mobile and immobile dislocation densities of RL111-1.

Between the shock front and the leading edge of the elastic–plastic release wave, the shear stress remains constant and dislocations are in an equilibrium state. During this period, no further dislocation evolution is observable, and mobile and immobile dislocation densities remain stable. Here, we assume that, despite their vanishing velocity, mobile dislocations are still mobile and will restart to move once RSS exceeds CRSS.

When the elastic release wave arrives, the stress continues decreasing reversely. Mobile dislocations restart to move once RSS exceeds CRSS. Dislocation evolution also resumes. As shown in Figure 8, the immobilization rate is one order of magnitude larger than the generation rate when the elastic wave arrives, which indicates that dislocation immobilization dominates the dislocation evolution during quasi-elastic release. Most of the mobile dislocations are immobilized quickly, which leads to a dramatic increase of immobile dislocation density, up to a peak value of ~$10^{14}$ m$^{-2}$ after plastic unloading ends (see the red curve in Figure 7), which is in accordance with experimental results.[20]

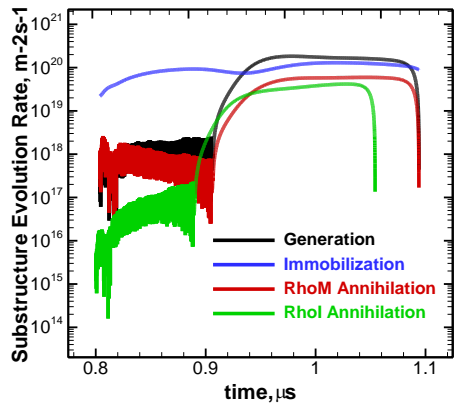

FIG. 8. Dislocation substructure evolution rate of one slip direction, along which the CRSS does not equal 0. The black, blue, red, and green lines denote the generation rate, the immobilization rate, the mobile dislocation annihilation rate, and the immobile dislocation annihilation rate, respectively.

Yield stress is proportional to the square root of the immobile dislocation density, as suggested by Equation (13). During quasi-elastic unloading, the above-mentioned rise in immobile dislocation density resulting from dislocation immobilization leads to a smooth increase of yield stress. As Figure 6 shows, RSS is always restricted to the level of CRSS and increases as CRSS increases during quasi-elastic release, which results in the particle velocity decreasing smoothly with the same tendency until the plastic release wave reaches the region. It is concluded that the dislocation immobilization during dynamic deformation causes the quasi-elastic release effect.

## A. Unloading speed

The above analysis shows that dislocation immobilization leads to quasi-elastic release. The immobilization rate is assumed to be proportional to the dislocation density with a characteristic velocity $V_I$. In the following, simulations of RL111-1 at different $V_I$ values are conducted to clarify its influence on unloading speed.

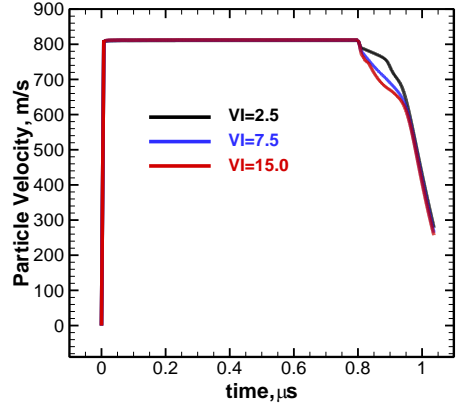

FIG. 9. Simulations of RL111-1 with different $V_I$ values.

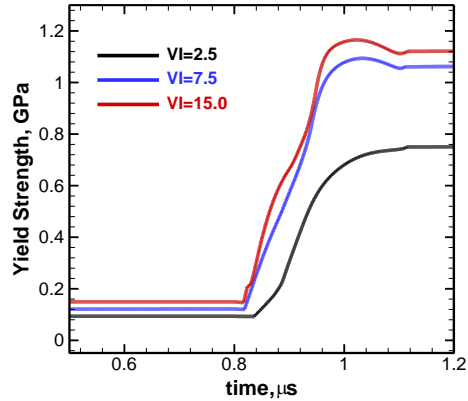

FIG. 10. *In situ* yield stress of RL111-1 for different $V_I$ values.

The calculated results at $V_I = 2.5$, $V_I = 7.5$, and $V_I = 15.0$ are presented in Figure 9. As shown in this figure, the unloading speed during quasi-elastic release increases as $V_I$ increases. It is inferred from the stress–velocity relation, namely, $\sigma = \rho D u$, that the evolution tendency of the particle velocity is similar to that of the stress. Figure 6 demonstrates that the stress is always relaxed to the level of the yield stress by dislocation evolution. Thus, the evolution tendency of the particle velocity should be similar to that of yield stress. The yield stress increases faster at higher $V_I$ owing to the higher immobilization rate caused by higher $V_I$, as shown in Figure 10. Therefore, the unloading speed during quasi-elastic release is positively related to $V_I$.

We can also learn from Figure 10 that the time derivative of the yield strength keeps decreasing until the end of quasi-elastic release, and this tendency is strengthened with higher $V_I$. Differentiating equation (15) with respect to time, one can get the time derivative of yield stress:

$$\frac{dY}{dt} \approx \frac{1}{2} A_I G b V_I \rho_D. \qquad (17)$$

Equation (17) suggests that the time derivative of yield stress is proportional to the mobile dislocation density. During quasi-elastic release, the mobile dislocation density decreases quickly, as does the time derivative of the yield stress, and this tendency is more significant at higher $V_I$, so does the time derivative of yield stress. It is expected that the unloading behavior will be closer to ideal elastic–plastic release when $V_I$ tends to infinity.

## B. The transition point

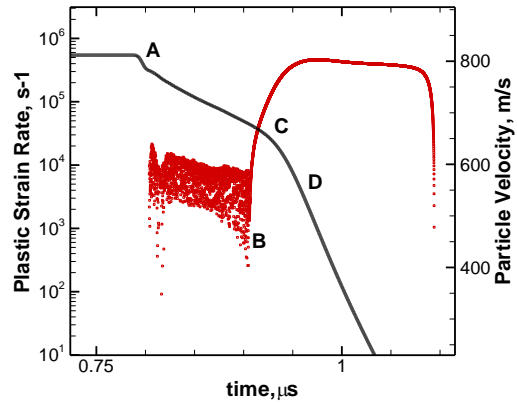

FIG. 11. Time history of the longitudinal plastic strain rate of RL111-1. The black line denotes the particle velocity and the red diamonds denote the plastic strain rate.

Besides the quasi-elastic phenomenon, another feature, namely, a transition point, is found between the quasi-elastic release and plastic release in the profile, as denoted by point *C* in Figure 11. After this point, the slope of the velocity profile is distinctly higher than that of the quasi-elastic release section. This point is commonly regarded as the critical point between elastic release and plastic release. We can also learn from Figures 11 and 12 that the plastic strain rate and the dislocation velocity oscillate with a small amplitude during quasi-elastic release.

Different physical processes lead to different features at different sections of the velocity profile. When the elastic release wave arrives, RSS is always slightly larger than CRSS and increases with increasing CRSS. The stress will be relaxed by dislocation slip, while the yield stress will increase as mobile dislocations become immobilized. During this process, the resultant force acting on dislocations, namely, RSS minus CRSS, oscillates with a small amplitude, which results in the oscillation of dislocation velocity and the plastic strain rate at the initial section of quasi-elastic release. After this part, the decrease of dislocation density leads to the decrease of plastic strain rate and the increase of dislocation velocity leads to the increase of the plastic strain rate until the plastic release wave arrives.

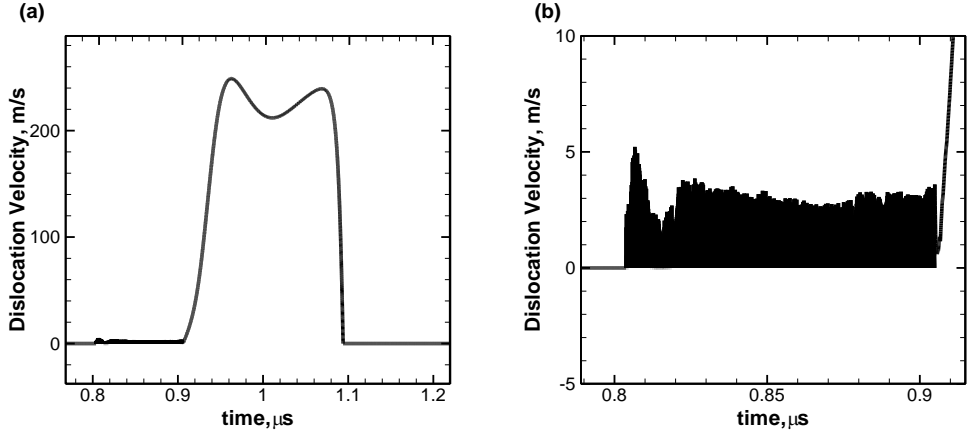

FIG. 12. Time history of the dislocation velocity in the {01-1}⟨111⟩ direction.

When the plastic release wave arrives, RSS begins to deviate from but is still close to CRSS, as shown in Figure 6(b), which leads to rapid generation of dislocations and a quick increase of the plastic strain rate, as denoted by the section between points *B* and *D*. We can learn from Figure 7 that point *B* also denotes the time when the mobile dislocation density reaches its minimum during quasi-elastic release. As the mobile dislocation density increases, the plastic strain rate increases to its maximum, which is several orders of magnitude larger than that at point *B*, as denoted by point *D*. The slope of the velocity profile is proportional to the strain rate at a lower level of deformation.[21] Therefore, the increase of the plastic strain rate between points *B* and *D* leads to the rapid increase of the slope of velocity profile, which results in the appearance of the transition point. After the increase of the plastic strain rate, the deformation rate becomes relatively stable until the end of the plastic release.

Summarizing the above analysis, we can state that the transient increase of the dislocation generation rate caused by the plastic release wave results in the appearance of the transition point.

# IV. SUMMARY AND CONCLUSIONS

A comprehensive study was undertaken to determine how the dislocation substructure and its evolution influence the dynamic deformation of shock-compressed aluminum. To accomplish this objective, grain-level heterogeneities are not considered during simulations and polycrystalline aluminum is treated as a single crystal since its anisotropy factor is near 1. The calculated results indicate that the competition between different dislocation evolution mechanisms, namely, dislocation generation and immobilization, leads to different features in the unloading profile. When the elastic release wave reaches a region, dislocation immobilization dominates dislocation evolution, which results in quasi-elastic release. Comparing the calculated results at different characteristic immobilization velocities indicates that the immobilization speed greatly influences the unloading speed. Dislocation generation becomes the main deformation mechanism when the plastic release wave arrives, which leads to the emergence of the transition point. As suggested by the above analysis, the time at which the plastic release wave arrives is appreciably earlier than the transition point, which indicates that features of the velocity profile lag behind the dynamic deformation in the material.

To obtain closer agreement with experimental data, the values of some parameters are obtained by fitting the experimental results. In particular, the generation coefficient *k* is much less than that in Ref. 16, in which this coefficient is proportional to the plastic deformation energy required for new dislocation generation. We did not offer a physical explanation for this difference, since the main objective of this work is to explain the quasi-elastic release effect with dislocation substructure evolution. Thus, additional work is required to clarify the physical mechanism that leads to this difference.

The quasi-elastic effect is found not only in unloading experiments but also in reloading experiments. Since the materials are in different states in these two experiments, the physical mechanisms that result in the quasi-elastic effect may be not the same. In particular, the material is in a compressive state in reloading experiments, whereas it is in a tensile state in unloading experiments. Therefore, a further study to distinguish the difference is necessary.

# ACKNOWLEDGEMENTS

The authors are grateful to Dr. Chen Shi for helping with the language and for useful remarks. This work was supported by the Foundation of the President of the China Academy of Engineering Physics (Grant No. 201402084), the National Natural Science Foundations of China (Grant Nos. 11302202, 11372294, and 11532012), the Foundation of the National Key Laboratory of Shock Wave and Detonation Physics (No. 9140C670301150C67290), and the Lingyu Guihua Project (Grant No. LYGH201502).